\documentclass[cits]{PoS}

\title{Abundances in Metal-Poor Stars and Chemical Evolution of the Early Galaxy}

\ShortTitle{Chemical Evolution of the Early Galaxy}

\author{\speaker{G. J. Wasserburg}%
         \thanks{Unable to attend the conference due to illness.}\\
        The Lunatic Asylum, Division of Geological and Planetary Sciences, California Institute of 
        Technology, Pasadena, CA 91125, USA\\
        E-mail: \email{gjw@gps.caltech.edu}}
        
\author{Y.-Z. Qian%
		\thanks{Spoke on behalf of G.J.W.}\\        
	School of Physics and Astronomy, University of Minnesota, Minneapolis, MN 55455, USA\\
        E-mail: \email{qian@physics.umn.edu}}

\abstract{We have attributed the elements from Sr through Ag in stars of low 
metallicities (${\rm [Fe/H]}\lesssim -1.5$) to charged-particle reactions
(CPR) in neutrino-driven winds, which are associated with neutron star 
formation in low-mass and normal supernovae (SNe) from progenitors 
of $\sim 8$--$11\,M_\odot$ and $\sim 12$--$25\,M_\odot$, respectively.
Using this rule and attributing all Fe production to normal SNe, we 
previously developed a phenomenological two-component model, which 
predicts that ${\rm [Sr/Fe]}\geq -0.32$ for all metal-poor stars. 
This is in direct conflict with the high-resolution data now available,
which show that there is a great shortfall of Sr relative to Fe in many 
stars with ${\rm [Fe/H]}\lesssim -3$. The same conflict also exists for 
the CPR elements Y and Zr. We show that the data require a stellar 
source leaving behind black holes and that hypernovae (HNe) from 
progenitors of $\sim 25$--$50\,M_\odot$ are the most plausible 
candidates. If we expand our previous model to include three components 
(low-mass and normal SNe and HNe), we find that essentially all of the 
data are very well described by the new model. The HN yield pattern for 
the low-$A$ elements from Na through Zn (including Fe) is inferred from
the stars deficient in Sr, Y, and Zr. We estimate that HNe contributed 
$\sim 24\%$ of the bulk solar Fe inventory while normal SNe contributed 
only $\sim 9\%$ (not the usually assumed $\sim 33\%$). This implies a 
greatly reduced role of normal SNe in the chemical evolution of the 
low-$A$ elements. This work was supported in part by US DOE grants 
DE-FG03-88ER13851 (G.J.W.) and DE-FG02-87ER40328 (Y.Z.Q.).
G.J.W. acknowledges NASA's Cosmochemistry Program for research support 
provided through J. Nuth at the Goddard Space Flight Center. He also
appreciates the generosity of the Epsilon Foundation.}

\FullConference{10th Symposium on Nuclei in the Cosmos\\
                 July 27 - August 1, 2008\\
                 Mackinac Island, Michigan, USA}

\begin{document}

\section{Introduction}
The high-resolution (e.g., \cite{johnson02,honda04,aoki05,francois07,cohen08} 
and medium-resolution \cite{barklem05} observations of elemental abundances 
in a large number of low-metallicity stars in the Galactic halo 
(and a single star in a dwarf galaxy \cite{fulbright04}) now provide 
a data base for determining the stellar sources contributing 
to the interstellar/intergalactic medium (ISM/IGM) at metallicities of 
$-5.5<{\rm [Fe/H]}<-1.5$. These data taken in conjunction with stellar models appear 
to define the massive stars active in the early epochs. This changes our views of 
what may be Population III (Pop III) stars and what stellar types are continuing 
contributors through the present epoch. A key to our understanding is the recognition 
that low-mass and normal core-collapse supernovae (SNe) from progenitors of
$\sim 8$--$11\,M_\odot$ and $\sim 12$--$25\,M_\odot$, respectively, end their lives 
as neutron stars and that nucleosynthesis in the neutrino-driven winds of nascent 
neutron stars produce a large group of elements including Sr, Y,  and Zr via 
charged-particle reactions (CPR, e.g., \cite{wh92}). This process is not the true 
``$r$-process'' with rapid neutron capture. In considering the contributions of 
low-mass and normal SNe that are certainly key contributors, certain rules are now 
established:

1. The true ``$r$-process'' is not connected with any significant production of Fe  group 
nuclei (or the low-$A$ elements below Zn with mass numbers $A\sim 23$--70, e.g., 
\cite{qw02});

2. The CPR nuclei are not directly coupled to the true ``$r$-process'' elements. While 
some workers consider that a ``turn-on'' of high neutron densities may be achieved in 
some unknown way coupled to a neutrino-driven wind, the observational data do not 
appear to be in support of this;

3. If one assumes that only low-mass and normal SNe were the sources, then [Sr/Fe] 
must exceed $-0.32$ for all metal-poor stars. 

\begin{figure}
\vskip -0.75cm
\begin{center}
\includegraphics[angle=270,width=0.5\textwidth]{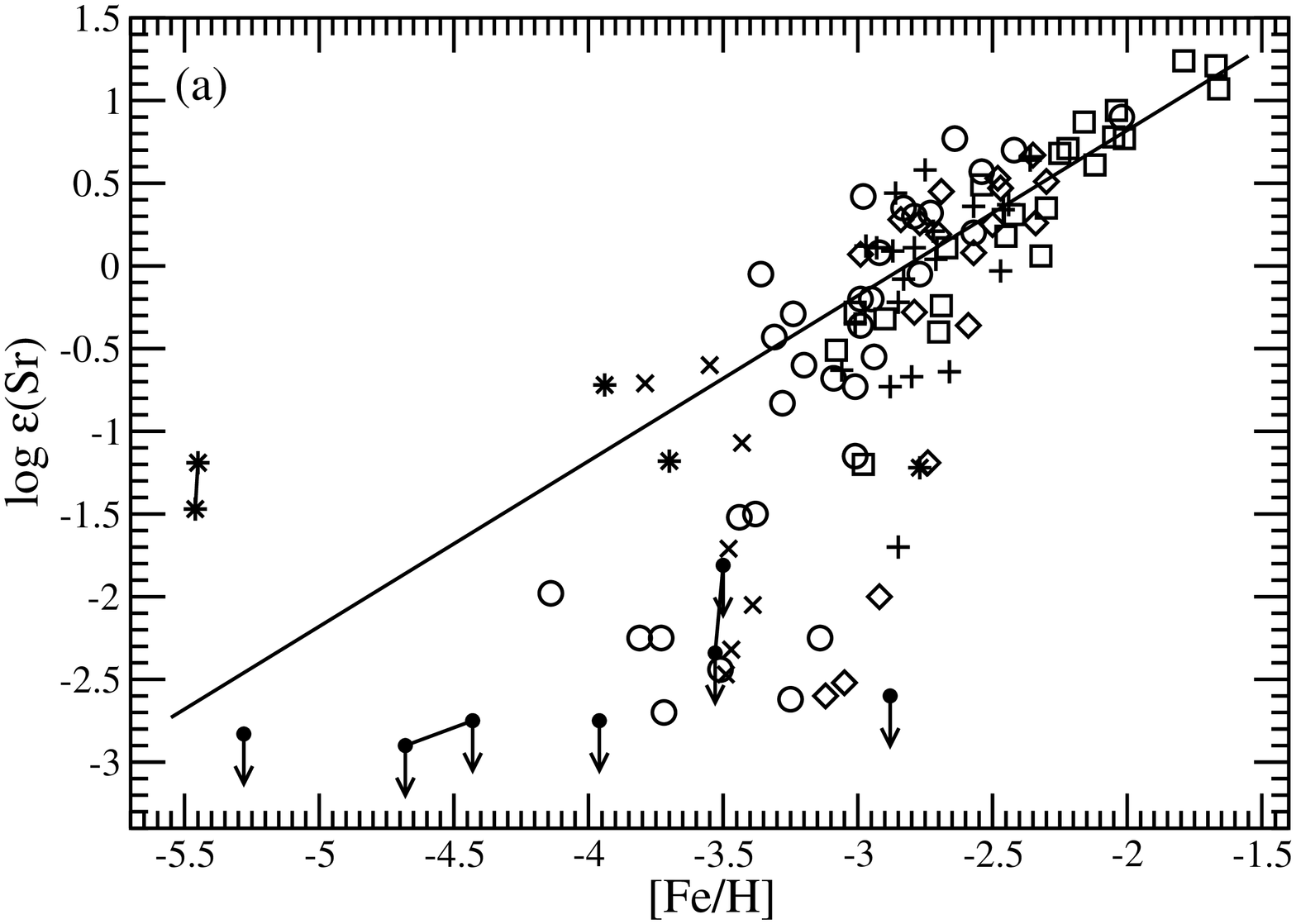}%
\includegraphics[angle=270,width=0.5\textwidth]{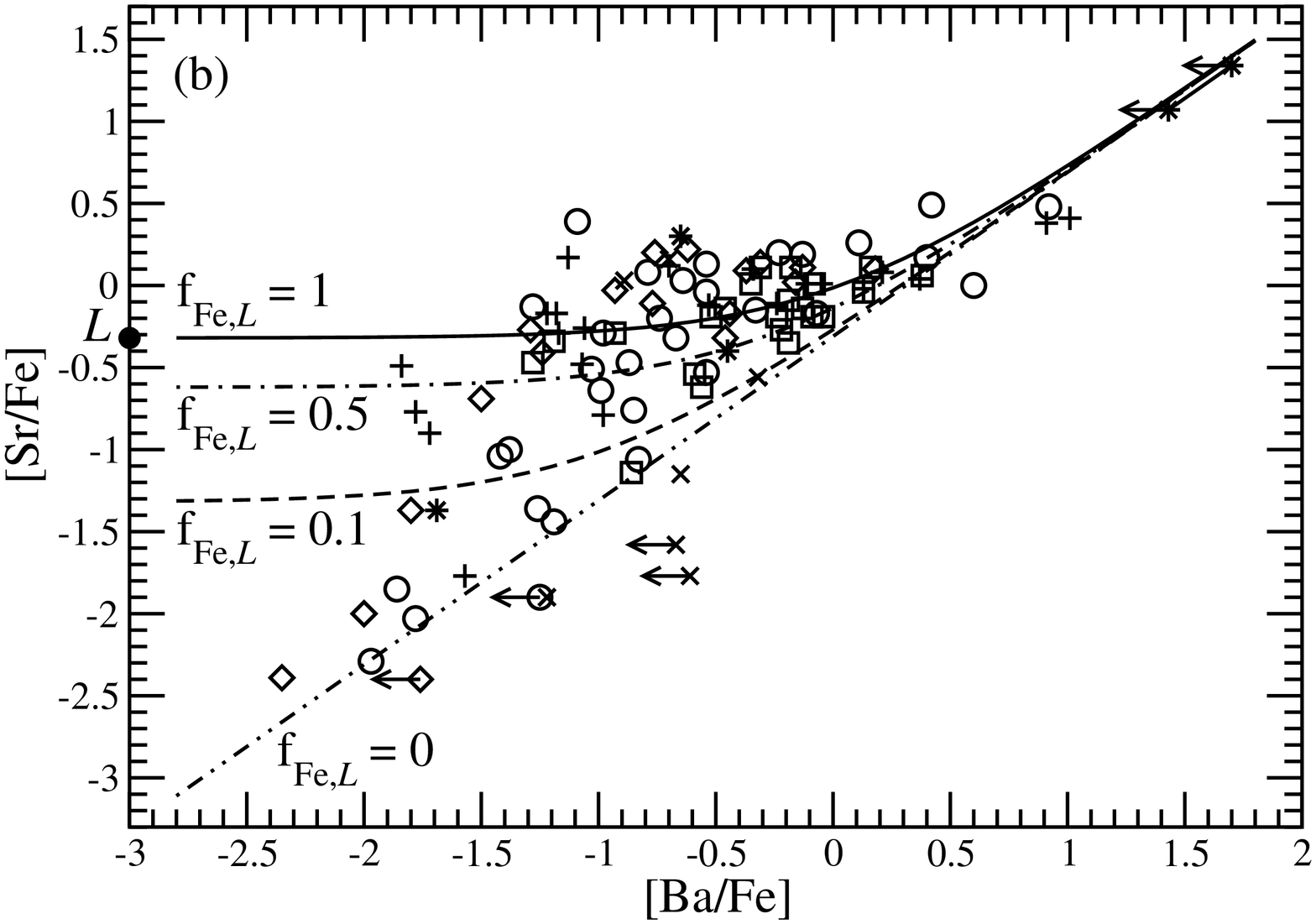}
\caption{(a) High-resolution data on $\log\epsilon({\rm Sr})$ vs. [Fe/H]. The solid line is 
calculated from the two-component model for a well-mixed ISM/IGM. Note that there is 
a great deficiency of Sr for many sample stars with ${\rm [Fe/H]}\lesssim-3$. 
It is evident that 
a source producing Fe and no Sr is required. (b) Evolution of [Sr/Fe] with [Ba/Fe] for
the data shown in (a). The curves correspond to different fractions $f_{{\rm Fe},L}$ of 
Fe due to the $L$ source (normal SNe). Details for these and the other figures of
this contribution can be found in \cite{qw08}.\label{fig1}}
\end{center}
\end{figure}

The last rule appears to be well followed for ${\rm [Fe/H]}>-3$, consistent with a 
two-component model of chemical evolution \cite{qw07}. However, below 
${\rm [Fe/H]}\sim -3$
there is a gross deficiency of Sr (and other CPR elements) relative to Fe. This is shown 
in Figure~\ref{fig1}a. This observation requires that there be an early stellar source 
of Fe that leaves behind a black hole instead of a neutron star with neutrino-driven winds
producing the CPR nuclei. It further follows that a third stellar component in addition to 
low-mass and normal SNe is required to account for all the abundance data.

\section{The third source}
The effects of all three stellar sources are most clearly seen if we consider the 
relationship of Sr (a CPR element) in conjunction with Ba (a true ``$r$-process'' 
element at low metallicities) and Fe. Using the yield ratios (Sr/Ba)$_H$ of the
$H$ source (low-mass SNe) and (Sr/Fe)$_L$ of the $L$ source (normal SNe)
we obtain: 
\begin{equation}
{\rm (Sr/H)}={\rm (Sr/Ba)}_H{\rm (Ba/H)}+{\rm (Sr/Fe)}_L{\rm (Fe/H)}f_{{\rm Fe},L},  
\end{equation}
where $f_{{\rm Fe},L}$ is the fraction of Fe from the $L$ source. The above 
equation can be rewritten as:
\begin{equation}
{\rm [Sr/Fe]}=\log\left(10^{{\rm [Sr/Ba]}_H+{\rm [Ba/Fe]}}+ f_{{\rm Fe},L}\times 
10^{{\rm [Sr/Fe]}_L}\right). 
\label{eq2}
\end{equation}
The evolution of [Sr/Fe] with [Ba/Fe] for the high-resolution data shown in 
Figure~\ref{fig1}a is exhibited in Figure~\ref{fig1}b along with the curves 
representing Equation~(\ref{eq2}) for $f_{{\rm Fe},L}=0$, 0.1, 0.5, and 1.
Similar results are found for the medium-resolution data. 

While there appears to be a clustering of data in the neighborhood of 
$f_{{\rm Fe},L}=1$ corresponding to Fe contributions exclusively from the $L$ 
source in Figure~\ref{fig1}b, there is a substantial fraction of the data lying 
down to $f_{{\rm Fe},L}=0$. This requires an Fe source not related to normal
SNe and clearly shows that the preponderance of the Fe in many sample 
stars is from this third source. Essentially the same results shown for [Sr/Fe]
vs. [Ba/Fe] are found for [Y/Fe] vs. [La/Fe] and [Zr/Fe] vs. [Ba/Fe], where La
and Ba are measures of the true ``$r$-process'' contributions. These results 
clearly show that there are major contributions from the third source producing 
Fe with no CPR elements and no ``$r$-process'' nuclei.

The matter at hand is what is the nature of this third source? Consideration of 
the yields of very massive stars ($\sim 140$--$260\,M_\odot$) associated with 
pair-instability SNe (PI-SNe) show that these sources are characterized by 
strong deficiencies in the elements of odd atomic numbers (e.g., Na, Al, K, Sc, 
V, Mn, Co). Neither the data from earlier studies \cite{mcw}
nor the more precisely-determined data from recent studies \cite{cayrel}
on low-metallicity halo stars exhibit these deficiencies. It follows that PI-SNe
cannot be the third source. A plausible candidate is hypernovae (HNe) from
progenitors of $\sim 25$--$50\,M_\odot$. These stars are known to be active 
in the current epoch, although little attention has been paid to them in 
consideration of Galactic chemical evolution. They have explosion energies 
far above those of low-mass and normal SNe and are presumed to be associated 
with gamma-ray bursts. The yields of HNe are generally not well known, but the 
typical Fe yield inferred from observations is $\sim 0.5\,M_\odot$,
much higher than the yield of $\sim 0.07\,M_\odot$ for normal SNe \cite{tominaga}.

\begin{figure}
\vskip -0.85cm
\begin{center}
\includegraphics[angle=270,width=0.6\textwidth]{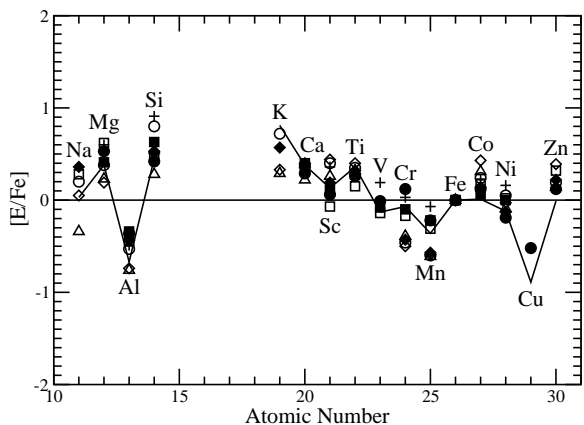}
\caption{The abundance patterns of the low-$A$ elements below Zn for stars with
``pure'' HN contributions ($f_{{\rm Fe},L}=0$ in Figure~\protect\ref{fig1}b). The solid 
curve represents the abundance pattern for a star with [Fe/H]~$=-2$.\label{fig2}}
\end{center}
\end{figure}

If we assume a Salpeter initial mass function (IMF) for massive stars of 
$\sim 8$--$50\,M_\odot$, the relative rates are $R_{\rm HN}:R_H:R_L
\sim 0.36:0.96:1$ for HNe, low-mass SNe ($H$), and normal SNe ($L$), respectively.
Of the Fe contributed by massive stars to the ISM/IGM, the fraction from HNe is
$\sim 0.72$. Thus HNe are the dominant Fe source at early epochs with contributions
far exceeding those of normal SNe. This seems to explain the earlier conundrum 
that the yield patterns for the low-$A$ elements below Zn in stars at very low 
metallicities ([Fe/H]~$<-3$) appear to be indistinguishable from what was attributed to 
normal SNe at higher metallicities (see Figure~\ref{fig2}) \cite{qw02}.

\begin{figure}
\vskip -0.75cm
\begin{center}
\includegraphics[angle=270,width=0.5\textwidth]{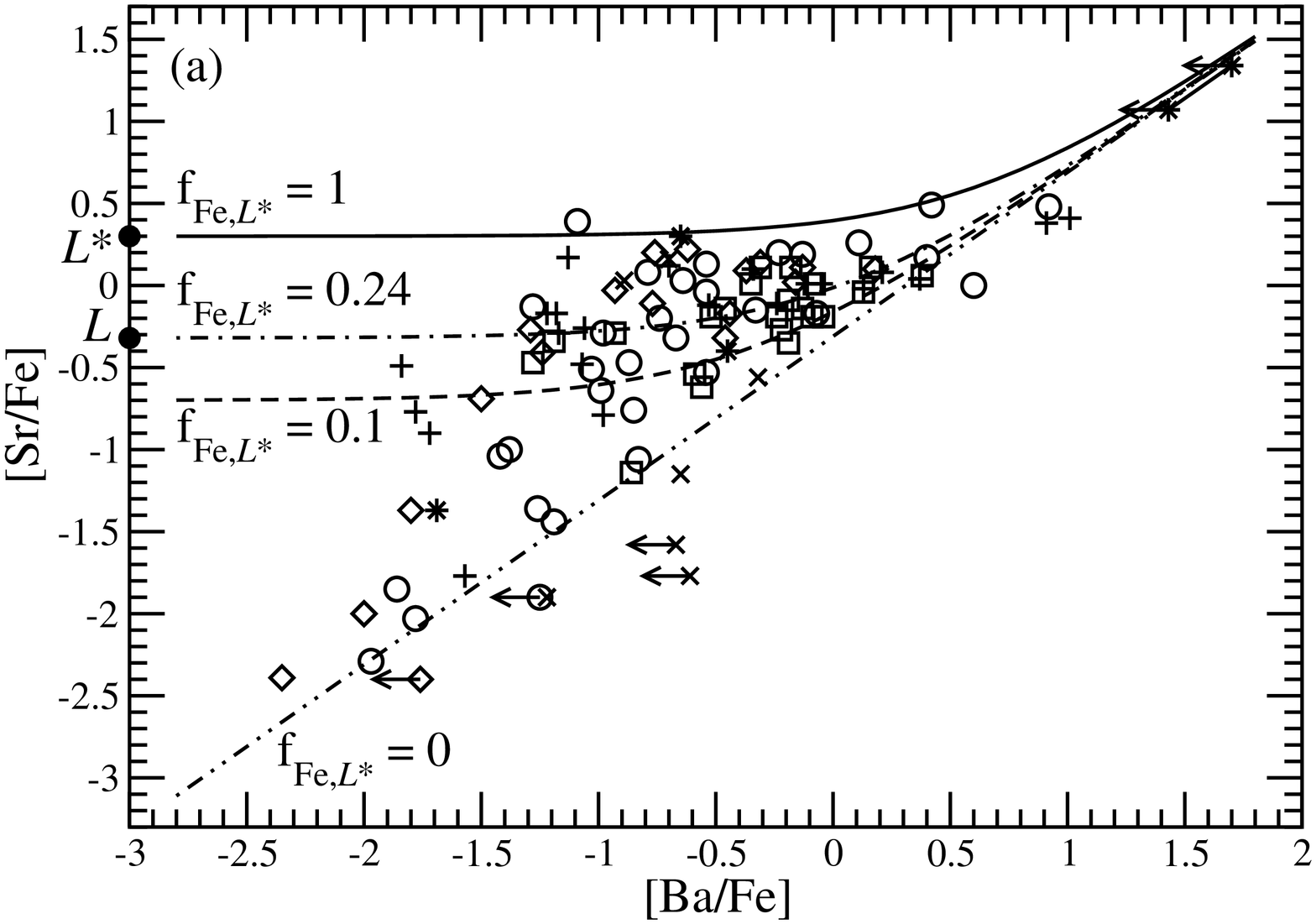}%
\includegraphics[angle=270,width=0.5\textwidth]{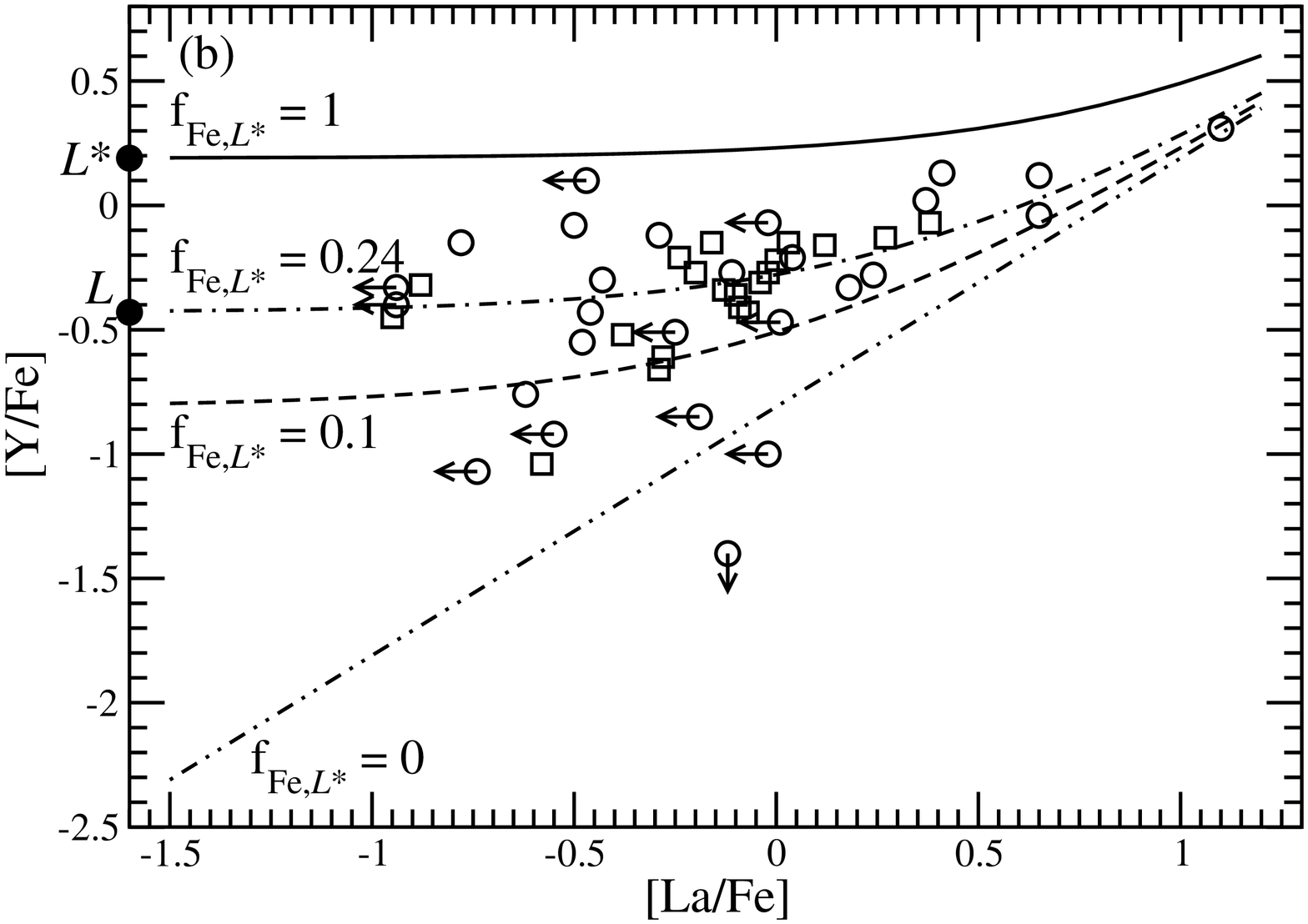}
\caption{(a) Similar to Figure~\protect\ref{fig1}b but with the $L$ source being
a blend of normal SNe ($L^*$) and HNe. (b) Same as (a) but for 
[Y/Fe] vs. [La/Fe].\label{fig3}}
\end{center}
\end{figure}

It is now evident that the inventory of the low-$A$ elements including Fe that we 
had attributed to normal SNe is in fact the mixture of HN and normal SN 
ejecta where the preponderant contributions are from HNe. Thus the 
previously-designated $L$ source is not a pure source but a blend of HNe and 
normal SNe: $L=L^*+\ {\rm HNe}$, where $L^*$ represents normal SNe. 
For the estimated yields and relative rates given above, it follows that the 
$\sim 1/3$ of the solar Fe inventory previously assigned to normal SNe is in 
considerable error. Taking $\sim 2/3$ of the solar Fe inventory to be from Type Ia
SNe,  we find that of the remaining $\sim 1/3$,  $\sim 24$\% is from HNe with 
only $\sim 9$\% from normal SNe. Thus with the proper assignment of Fe 
contributions for the $L$ blend with [Sr/Fe]$_L=-0.32$, we obtain 
[Sr/Fe]$_{L^*}=0.3$ for the $L^*$ source. Using the proper yield ratios for the 
$L^*$ source, we obtain the results shown in Figure~\ref{fig3}. 
It is seen that the data for all the elements appear to be described very well
by the three-component ($H$, $L^*$, and HNe) model. We further note that 
the clump of data above the $f_{{\rm Fe},L} =1$ curve in Figure~\ref{fig1}b 
are now absent in Figure~\ref{fig3}a with diminished Fe contribution from 
normal SNe.

\section{General discussion}
From the results reviewed above it appears that the whole chemical evolution 
in the ``juvenile epoch'' of the first Gyr after the Big Bang may be explained by 
the concurrent contributions of massive stars associated with low-mass and
normal SNe and HNe in a standard IMF and that this same relative contribution 
continues into the present epoch. The efforts to seek Pop III stars that only occur 
in early epochs and then stop are considered by us to be invalid as were our 
earlier efforts to find a ``prompt inventory'' in the ISM/IGM. It follows that models
for the formation of the ``first'' stars, which has been the focus of intensive studies 
with due consideration of the complex condensation and cooling processes at 
zero to low metallicities (e.g., \cite{abel02,bromm04}), should consider the stellar 
populations inferred here with HNe ($\sim 25$--$50\,M_\odot$) being the dominant 
metal source. This source is
highly disruptive and certainly can disperse debris through the IGM until halos of 
substantial mass have formed. The apparent 
sudden onset of heavy ``$r$-process'' elements, which motivated our earlier 
search for a ``prompt inventory'', is most plausibly related to the formation of 
halos of sufficient mass that remain bound following both SNe and HNe \cite{qw04}. 
It also follows that the earlier models of Galactic chemical evolution that aimed to
provide $\sim 1/3$ of the solar Fe inventory by normal SNe must now be subject to 
reinvestigation. The observational evidence for ongoing HNe in the current epoch 
cannot be ignored. There is further the fact that production of heavy (true) 
$r$-process nuclei is strongly decoupled from Fe production. The extent to which 
this presents a further challenge to stellar models is to be resolved.

\end{document}